\documentclass[preprint,aps]{revtex4}
\usepackage{graphicx}
\usepackage{amssymb}
\newcommand{\be}{\begin{eqnarray}}
\newcommand{\ee}{\end{eqnarray}}
\usepackage{hyperref}

\begin{document}

\title{Generalized Parton Distributions of the Photon}

\author{\bf Asmita Mukherjee and Sreeraj Nair}

\affiliation{ Department of Physics,
Indian Institute of Technology Bombay,\\ Powai, Mumbai 400076,
India.}
\date{\today}

\begin{abstract}
We present a first calculation of the generalized parton distributions of the
photon (both polarized and unpolarized) using overlaps of light-front wave
functions at leading order in $\alpha$ and zeroth order in $\alpha_s$; for 
non-zero transverse momentum transfer and zero skewness. We present the 
novel parton content of the photon in transverse position space.  
\end{abstract}

\maketitle

%%%%%%%%%%%%%%%%%%%%%%%%%%%%%%%%%%%%%%%%%%%%%%%%%%%%%%%%%%%%%%%%%%%%
{\bf{Introduction}}
\vspace{0.2in}

In deep inelastic scattering of a highly virtual photon on a real photon,
the partonic constituents of the photon play a dominant role when the
virtuality $Q^2$ is very large. In this case the pointlike contribution to
the photon structure function  $F_2^\gamma(x,Q^2)$ dominates over the hadronic
component. This can be calculated perturbatively. These are relevant in
the context of  $e^+e^-$
annihilation and photoproduction. Unlike the proton
structure function, where only the $Q^2$ dependence is calculated
perturbatively and the $x$ dependence has to be fitted using experimental
data, in the photon structure function, both $x$ and $Q^2$ dependence can be
calculated. $F_2^\gamma(x,Q^2)$ show logarithmic $Q^2$ dependence already in
parton model, unlike the proton structure function \cite{zerwas}. Leading order QCD
calculation differs from the parton model result for $F_2^\gamma(x,Q^2)$
by calculable finite terms \cite{witten}. The photon structure function is now known 
fairly accurately and agrees well with experimental results \cite{buras}.     

In \cite{pire} deeply virtual Compton scattering (DVCS) $\gamma^* \gamma \rightarrow
\gamma \gamma$ on a photon target was considered in the kinematic region of
large center-of-mass energy, large virtuality $(Q^2)$ but small squared
momentum transfer $(-t)$. The result was interpreted at leading logarithmic
order as a factorized form of
the scattering amplitude in terms of a hard handbag diagram and the
generalized parton distributions of the photon. The calculation was done at
leading order in $\alpha$ and zeroth order in $\alpha_s$
when the momentum transfer was purely in the longitudinal direction. 
These are called anomalous GPDs as they show logarithmic
scale dependence already in parton model. They are of particular interest as
they can be calculated in perturbation theory and can act as theoretical tools 
to understand the basic properties of GPDs like polynomiality and
positivity. Beyond leading logarithmic order one would need to include the
non-pointlike hadronic contribution. Note that the same process in a different
 kinematic region,
namely at low energy and high squared momentum transfer $-t \approx Q^2$,
gives information on the generalized distribution amplitudes (GDA) of the
photon, which describes the coupling of a quark-antiquark pair to the two
photons and are connected to the photon GPDs by crossing \cite{GDA}. 
DVCS process on a proton  target has been analyzed in
detail  theoretically and are also being accessed in experiments \cite{rev}. 
The proton GPDs are richer in content than the ordinary parton distributions (pdfs).
In the forward limit of zero momentum transfer they reduce to pdfs and their
$x$ moments give nucleon form factors. An interesting physical 
interpretation of GPDs has been obtained in \cite{burkardt} by taking their
Fourier transform with respect to the transverse momentum transfer. 
When the longitudinal momentum transfer is zero, this gives the distribution
of partons in the nucleon in the transverse plane. They are called impact 
parameter dependent parton distributions (ipdpdfs) $q(x,b^\perp)$. In fact
they obey certain positivity constraints which justify their physical
interpretation as probability densities. This interpretation holds in the
infinite momentum frame (even the forward pdfs have a probabilistic
interpretation only in this frame)  and there is no relativistic correction
to this identification because in light-front formalism, as well as 
in the infinite momentum frame, the transverse boosts act like
non-relativistic Galilean boosts.  When the nucleon is transversely polarized, 
the unpolarized impact parameter dependent pdf is distorted in the
transverse plane. A combination of chiral odd GPDs in impact parameter space
gives information on the correlation between the spin and orbital angular
momentum of the quarks inside the target \cite{chiral}. Fourier transform 
(FT) with respect to the skewness $\zeta$ gives rise to a diffraction-like pattern
\cite{hadron_optics}.  Thus the GPDs in effect give a complete 
(Lorentz invariant) 3 D 
picture of the proton in position space. While the proton is known to be a 
composite particle, it is interesting to access the partonic structure of 
the photon probed in high energy processes. As the proton GPDs are richer in
content than the ordinary pdfs, photon GPDs can shed more light on the
partonic content of the photon.

In this letter, we calculate the photon GPDs using overlaps of light-front wave
functions. We take the momentum transfer to be purely in the transverse
direction, unlike \cite{pire}, where the momentum transfer was taken purely in the
lightcone (plus) direction. We keep leading logarithmic terms and
the mass terms coming from the vertex; and upto leading order in
electromagnetic coupling and zeroth order in strong coupling. 
There are contributions only from the diagonal
(particle number conserving) overlaps. When there is nonzero momentum
transfer in the longitudinal direction, there are off-diagonal particle
number changing overlaps as well, similar to the proton GPDs \cite{overlap}. 
The two particle light-front wave
functions of the photon can be calculated analytically using perturbation
theory. Taking a Fourier
transform with respect to the momentum transfered in the transverse
direction,  $\Delta^\perp$,  we express the GPDs in the
transverse impact parameter space. 

\vspace{0.2in}
      
%%%%%%%%%%%%%%%%%%%%%%%%%%%%%%%%%%%%%%%%%%%%%%%%%%%%%%%%%%%%%%%%%%%%%%%%%%%%
{\bf{GPDs of the photon}}
%%%%%%%%%%%%%%%%%%%%%%%%%%%%%%%%%%%%%%%%%%%%%%%%%%%%%%%%%%%%%%%%%%%%%%%%%%%%
\vspace{0.2in}

The GPDs for the photon can be expressed as the following off-forward matrix
elements defined for the real photon (target) state \cite{pire}:

\be
F^q=\int {dy^-\over 8 \pi} e^{-i P^+ y^-\over 2} \langle \gamma(P') \mid
{\bar{\psi}} (0) \gamma^+ \psi(y^-) \mid \gamma (P)\rangle ;
\nonumber\\ 
\tilde F^q=\int {dy^-\over 8 \pi} e^{-i P^+ y^-\over 2} \langle \gamma(P') \mid
{\bar{\psi}} (0) \gamma^+ \gamma^5 \psi(y^-) \mid \gamma (P)\rangle .
\ee
$F^q$ contributes when the photon is unpolarized and $\tilde F^q$ is the
contribution from the polarized photon. We have chosen the light-front gauge
$A^+=0$. 
As pointed out in \cite{pire}, there is also the photon operator which mixes with the
quark operator, it contributes at the same order in $\alpha$ in the
scattering amplitude. However here we calculate the matrix element rather
than the amplitude of the process and the contribution to the matrix element
of the photon operator comes at zeroth order in $\alpha$.
$F^q$ and $\tilde F^q$ can be calculated using the Fock space expansion of
the state, which can be written as \cite{pire}
\be
\mid \gamma(P)\rangle &=& \sqrt{N} \Big [ a^\dagger(P, \lambda) \mid 0
\rangle + \sum_{\sigma_1, \sigma_2} \int \{dk_1\} \int \{ dk_2\} \sqrt{2{(2
\pi)}^3 P^+} \delta^3 (P-k_1-k_2)\nonumber\\&&~~~~ \phi_2(k_1,k_2,\sigma_1, \sigma_2)
b^\dagger(k_1, \sigma_1) d^\dagger(k_2, \sigma_2) \mid 0 \rangle \Big ]
\ee 
where $\sqrt{N}$ is the overall normalization of the state; which in our
calculation we can take as unity as any correction to it contributes at
higher order in $\alpha$.  $\{ dk\}= \int {dk^+ d^2 k^\perp\over \sqrt{2 {(2
\Pi)}^3 k^+}}$, $\phi_2$ is the
two-particle ($q \bar{q}$) light-front wave function (LFWF) and $\sigma_1$ and
$\sigma_2$ are the helicities of the quark and antiquark. The wave function
can be expressed in terms of Jacobi momenta $x_i={k_i^+\over P^+}$ and
$q_i^\perp=k_i^\perp-x_i P^\perp$. These obey the relations $\sum_i x_i=1,
\sum_i q_i^\perp=0$. The boost invariant LFWFs are given by     
${\psi_2(x_i,q_i^\perp)=\phi_2 \sqrt{P^+}}$. $\psi_2(x_i, q_i^\perp)$
can be calculated order by order in perturbation theory. The two-particle
LFWFs are given by \cite{kundu}
\be
\psi_{2 s_1, s_2}^\lambda(x,q^\perp) &=& {1\over m^2-{m^2+{(q^\perp)}^2
\over x (1-x)}} {e e_q\over \sqrt{ 2 {(2 \pi)}^3}} 
\chi^\dagger_{s_1} \Big [
{(\sigma^\perp \cdot q^\perp)\over x} \sigma^\perp \nonumber\\&&- \sigma^\perp
{(\sigma^\perp \cdot q^\perp)\over 1-x} -i {m \over x (1-x)} \sigma^\perp
\Big ] \chi_{-s_2} \epsilon^{\perp *}_{\lambda} 
\ee
where we have used the two-component formalism \cite{two,kundu} and $m$ 
is the mass of
$q(\bar{q})$. $\lambda$ is the helicity of the photon and $s_1,s_2$ are the
helicities of the $q$ and ${\bar q}$ respectively. There is no contribution
to the matrix elements that we consider from the single particle sector of
the Fock space expansion. The leading term is the two-particle contribution,
which can be written as,
\be 
F^q &=& \int d^2 q^\perp dx_1 \delta(x-x_1) \psi_2^*(x_1,q^\perp-(1-x_1) 
\Delta^\perp
)\psi_2(x_1, q^\perp)\nonumber\\&&~~- \int d^2 q^\perp dx_1 \delta(1+x-x_1) 
\psi_2^*(x_1,q^\perp+x_1 \Delta^\perp
)\psi_2(x_1, q^\perp) 
\ee
Here we have suppressed the helicity indices and the sum over them. The
momentum transfered square $t$ is given by $t=(P-P')^2=-{(\Delta^\perp)}^2$. The first
 term is the contribution from the quarks and the second is the
contribution from the antiquark in the photon. As the light-cone momentum
fraction $x_1$ has to be always greater than zero, the first term
contributes when $1>x>0$ and the second term for $-1<x<0$. Using the LFWFs
each component can be calculated separately. We calculate in the same
reference frame as \cite{hadron_optics}. Note that the light cone plus
momentum of the target photon is non-zero. Finally we get for the unpolarized
photon
\be
F^q &=& \sum_q {\alpha e_q^2 \over 4 {\pi}^2 } \Big [ ((1-x)^2+x^2) (I_1+I_2+L
I_3) +2 m^2 I_3 \big ] \theta(x) \theta(1-x) 
\nonumber\\&&~~-\sum_q {\alpha e_q^2 \over 4 {\pi}^2 } 
\Big [ ((1+x)^2+x^2) (I'_1+I'_2+ L'
I'_3) +2 m^2 I'_3 \Big ] \theta(-x) \theta(1+x)
\ee
Here the sum indicates sum over different quark flavors; $L=-2 m^2 +2 m^2 x (1-x)
 - {(\Delta^\perp)}^2 (1-x)^2$, 
$L'=-2 m^2 -2 m^2 x (1+x) - {(\Delta^\perp)}^2 (1+x)^2$; the integrals can be
written as,
\be
I_1=\int {d^2 q^\perp \over D} = \pi Log \Big [{\Lambda^2\over \mu^2-m^2 x (1-x)
+m^2}\Big ]=I_2\nonumber\\
I_3= \int {d^2 q^\perp\over D D'}= \int_0^1 d \alpha {\pi\over P(x, \alpha,
 {(\Delta^\perp)}^2)}
\ee where $D={(q^\perp)}^2 -m^2 x (1-x) +m^2$ and $D'= {(q^\perp)}^2
+{(\Delta^\perp)}^2 (1-x)^2 -2 q^\perp \cdot \Delta^\perp (1-x) -m^2 x (1-x)
+m^2$, and $P(x, \alpha, {(\Delta^\perp)}^2)= -m^2 x (1-x) +m^2 + \alpha(1-\alpha)
 (1-x)^2 {(\Delta^\perp)}^2$. At zeroth order in $\alpha_s$ the
results are scale dependent, this scale dependence in our approach comes
from the upper limit of the transverse momentum integration $ \Lambda = Q$.
$\mu$ is a lower cutoff on the transverse momentum, which can be taken to
zero as long as the quark mass is nonzero. Leading order evolution of
the photon GPDs has been calculated in \cite{pire} for non-zero $\zeta$.
The mass terms in the vertex give subdominant contributions which we included.  

\vspace{3mm}

\begin{figure}[!htp]
%\centering
\begin{minipage}[c]{0.9\textwidth}

\tiny{(a)}\includegraphics[width=9.5cm,height=7cm,clip]{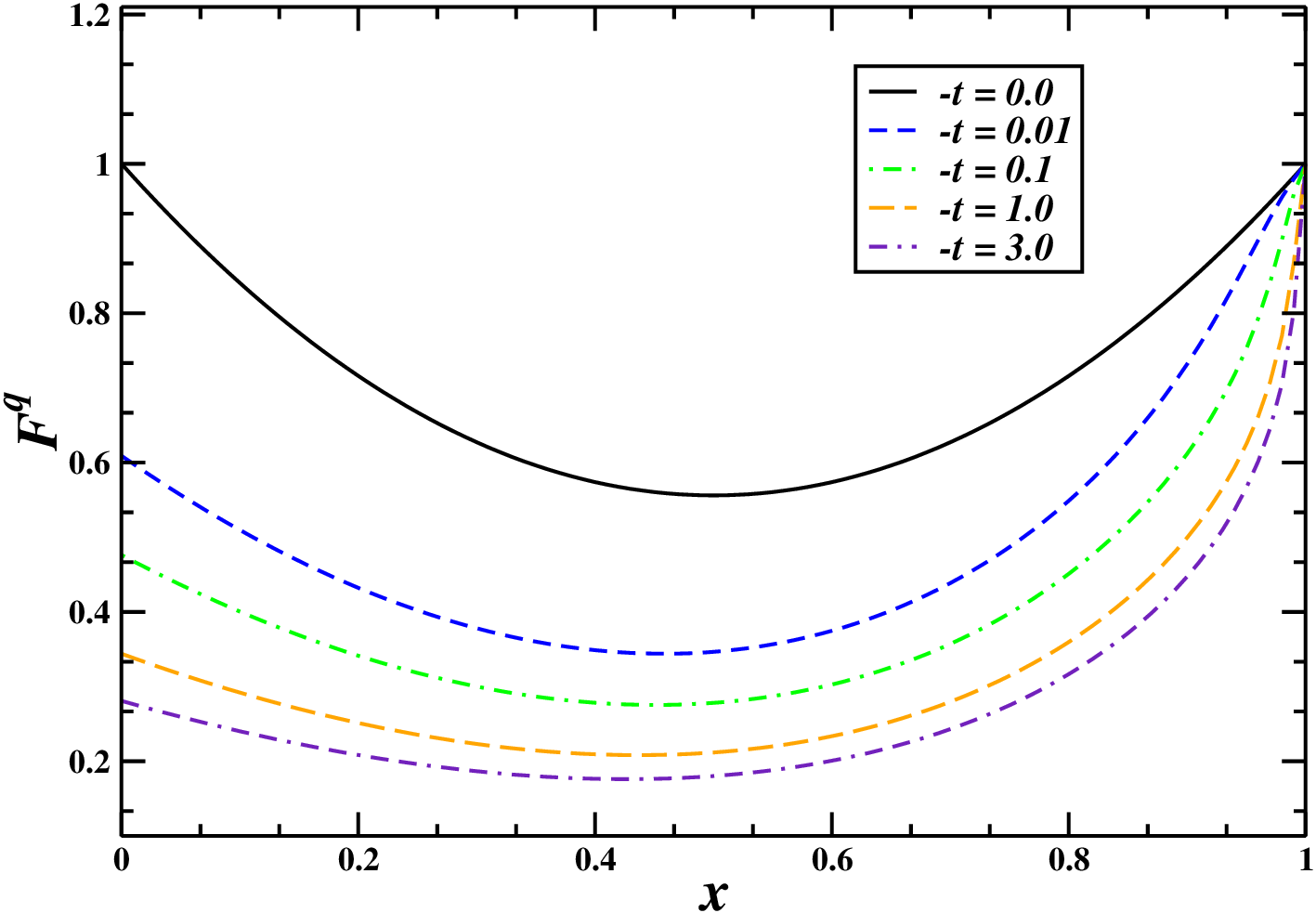}

\vspace{0.3in}

\tiny{(b)}\includegraphics[width=9.5cm,height=7cm,clip]{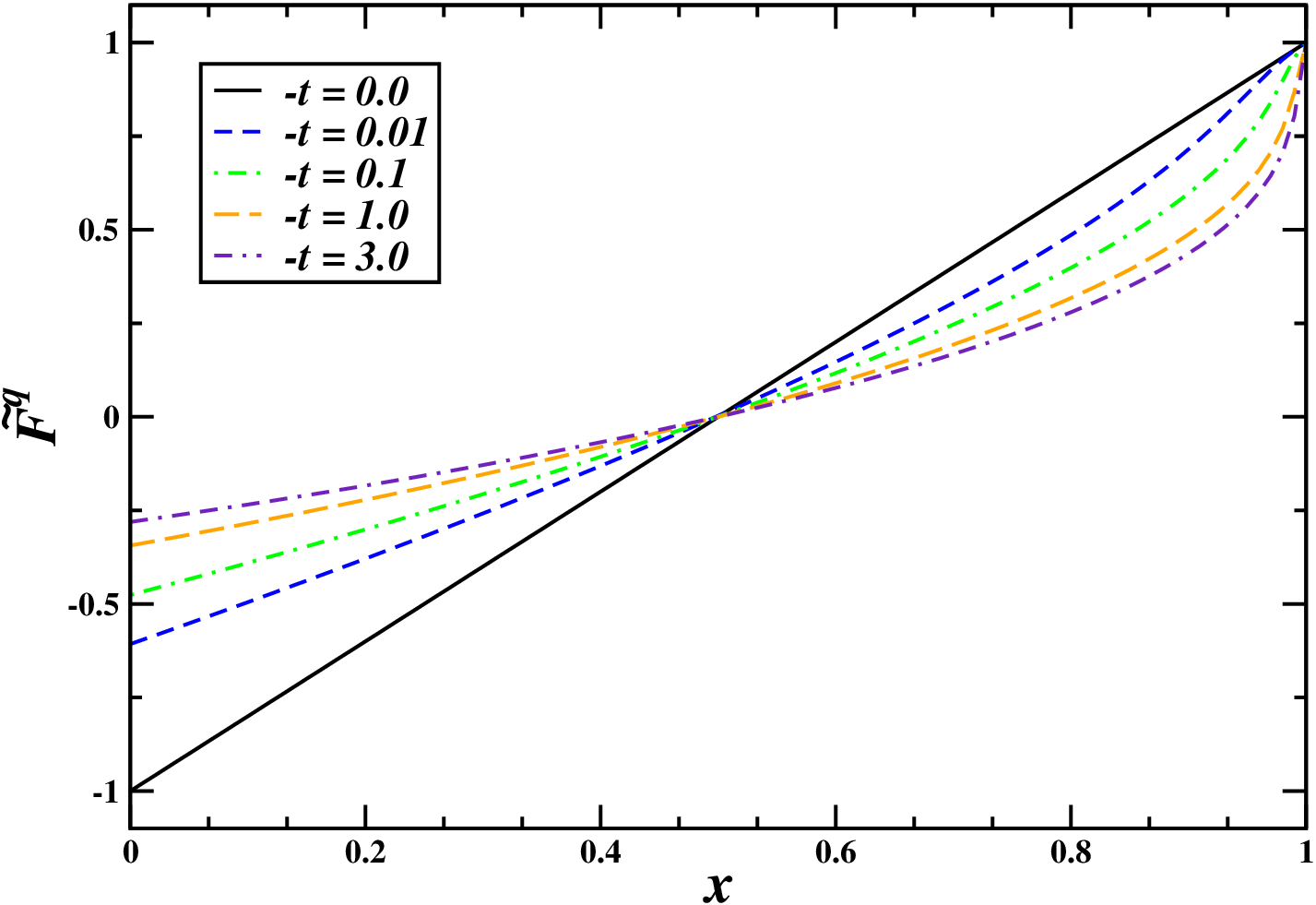}

\end{minipage}
\caption{\label{fig1}(Color online) (a) Plot of unpolarized
GPD $F^{q}$ vs $x$ for fixed values of $-t$ in $GeV^{2}$
and 
(b) polarized GPD  $\tilde{F^{q}}$ vs $x$ for fixed values of $-t$ in
$GeV^{2}$, $\Lambda = 20 \mathrm{GeV}$. In both plots the
normalization factor is chosen to compare with \cite{pire} when $t=0$.}

\end{figure}
\vspace{4mm}

For the antiquark contributions we have similar integrals
\be
I'_1=\int {d^2 q^\perp \over H} = \pi Log \Big [{\Lambda^2\over \mu^2
+m^2 x(1+x)+m^2}\Big ]=I'_2\nonumber\\
I'_3= \int {d^2 q^\perp\over H H'}= \int_0^1 d \alpha {\pi\over Q(x, \alpha,
 {(\Delta^\perp)}^2)}
\ee  
where $H={(q^\perp)}^2 +m^2 x (1+x) +m^2$ and $H'= {(q^\perp)}^2
+{(\Delta^\perp)}^2 (1+x)^2 +2 q^\perp \cdot \Delta^\perp (1+x) +m^2 x (1+x)
+m^2$, and $Q(x, \alpha, {(\Delta^\perp)}^2)= m^2 x (1+x) +m^2 
+ \alpha(1-\alpha) (1+x)^2 {(\Delta^\perp)}^2$.

\begin{figure}[!htp]
%\centering
\begin{minipage}[c]{0.9\textwidth}

\tiny{(a)}\includegraphics[width=9.5cm,height=7cm,clip]{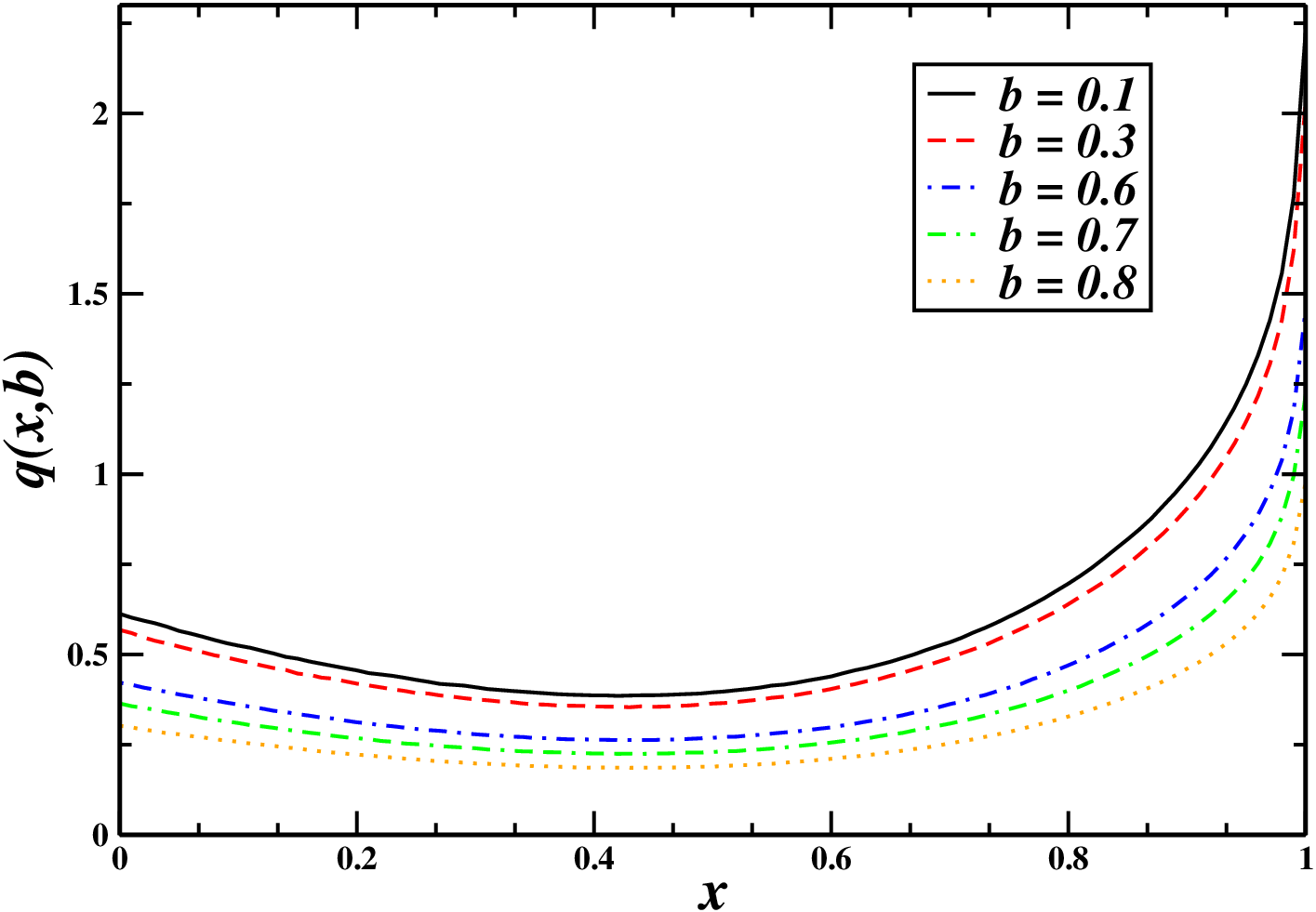}\\
\vspace{0.3in}
\tiny{(b)}\includegraphics[width=9.5cm,height=7cm,clip]{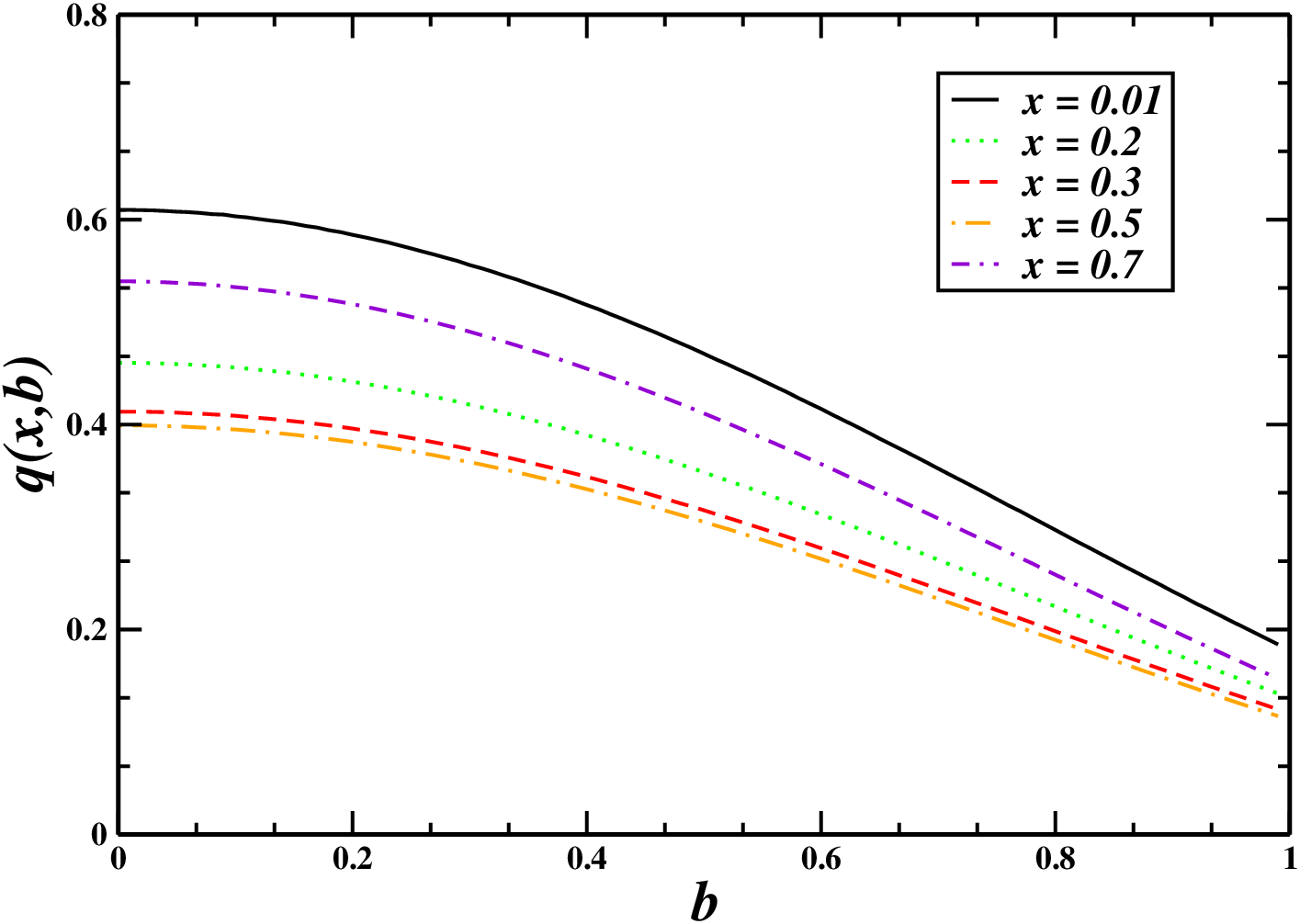}\\
\end{minipage}

\caption{\label{fig2}(Color online) (a)Plot of impact parameter dependent
pdf $q(x,b)$ vs $x$  for fixed $b$ values and (b) $q(x,b)$ vs $b$ for fixed
values of $x$
where we have taken $\Lambda$ = 20 $\mathrm{GeV}$ and $\Delta_{max}$= 3
GeV where $\Delta_{max}$ is the upper limit in
the $\Delta$ integration. $b$ is in ${\mathrm{GeV}}^{-1}$ and  $q(x,b)$ 
is in ${\mathrm{GeV}}^{2}$ .}
\end{figure}

For polarized photon the GPD $\tilde F^q$ can be calculated from the 
terms of the form 
$\epsilon^2_\lambda \epsilon^{1*}_\lambda-\epsilon^1_\lambda
\epsilon^{2*}_\lambda$ \cite{pire}. We consider the terms where the photon
helicity is not flipped. This can be written as, 
\be
\tilde F^q &=& \sum_q {\alpha e_q^2 \over 4 {\pi}^2 } \Big [ (x^2-(1-x)^2) (I_1+I_2+L
I_3) +2 m^2 I_3 \big ] \theta(x) \theta(1-x) 
\nonumber\\&&+\sum_q {\alpha e_q^2 \over 4 {\pi}^2 } \Big [ 
(x^2-(1+x)^2) (I'_1+I'_2+ L'
I'_3) +2 m^2 I'_3 \Big ] \theta(-x) \theta(1+x)
\ee

\begin{figure}[!htp]
%\centering
\begin{minipage}[c]{0.9\textwidth}

\tiny{(a)}\includegraphics[width=9.5cm,height=7cm,clip]{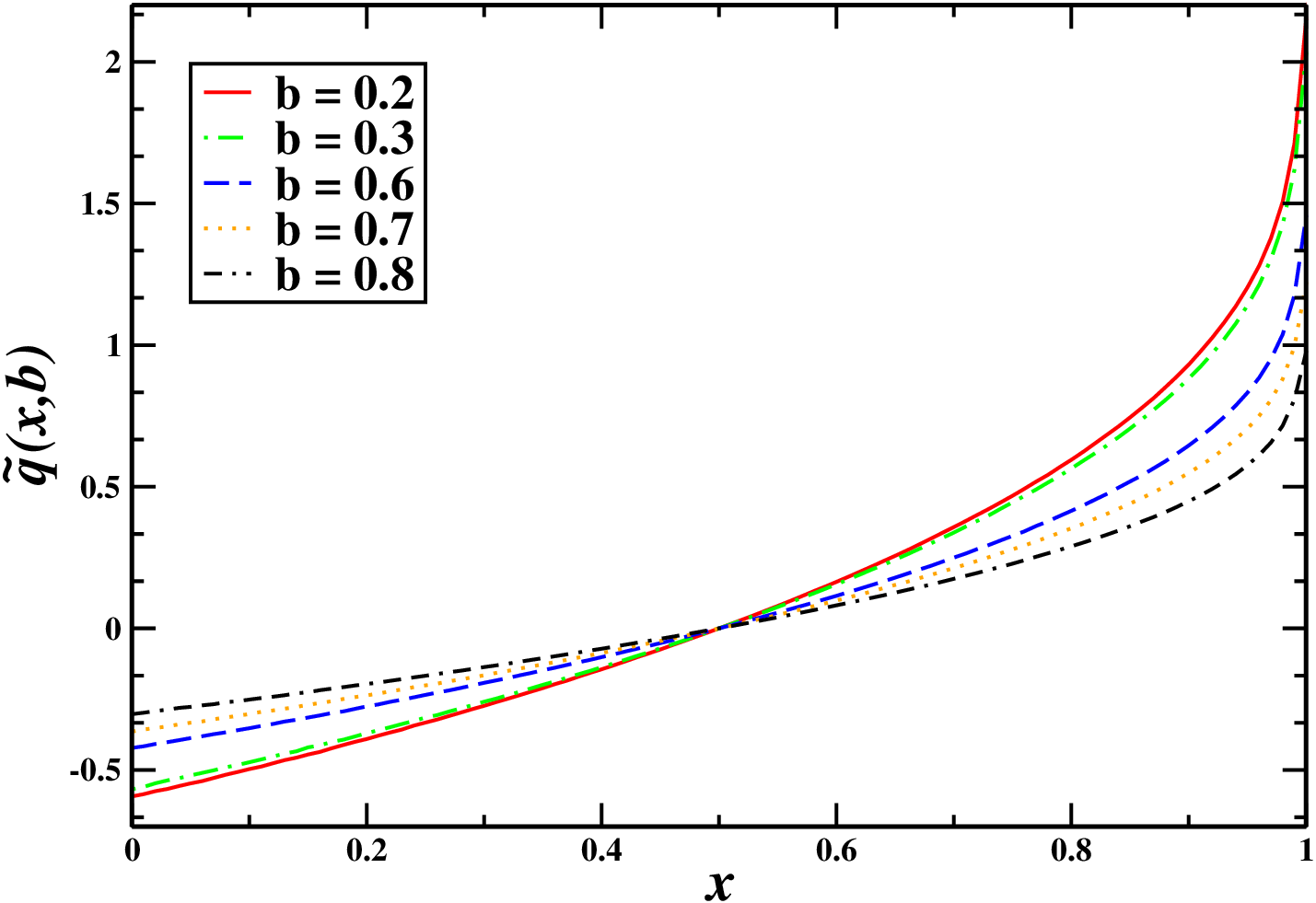}\\
\vspace{0.3in}

\tiny{(b)}\includegraphics[width=9.5cm,height=7cm,clip]{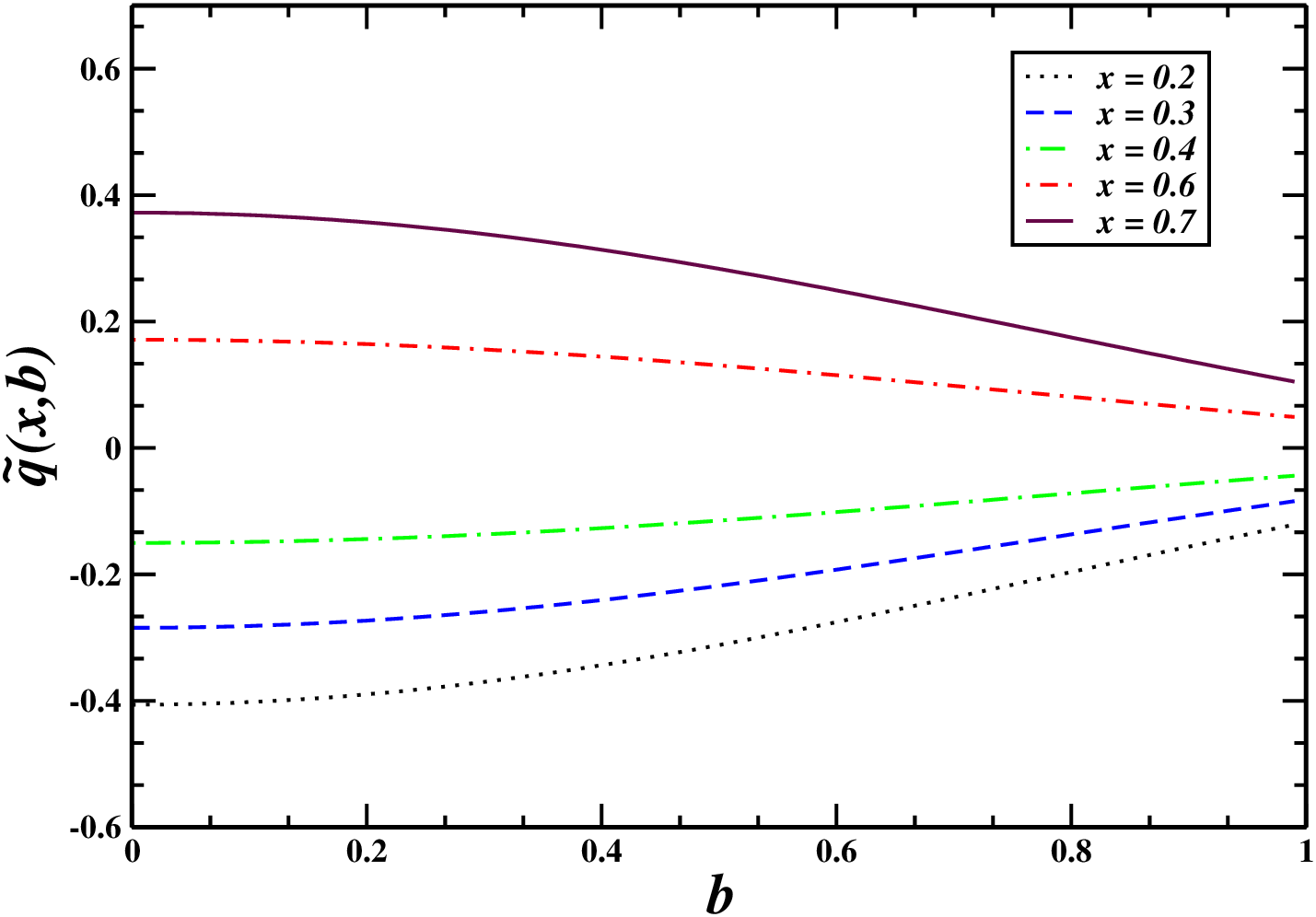}\\
\end{minipage}

\caption{\label{fig3}(Color online) (a)Plot of impact parameter dependent
pdfs $\tilde{q}(x,b)$  vs $x$  for fixed $b$ values and (b) $\tilde{q}(x,b)$ vs
$b$ for fixed values of $x$
where we have taken $\Lambda$ = 20 $\mathrm{GeV}$ and $\Delta_{max}$= 3
GeV where $\Delta_{max}$ is the upper limit in
the $\Delta$ integration. $b$ is in ${\mathrm{GeV}}^{-1}$ and $\tilde{q}(x,b)$ 
is in ${\mathrm{GeV}}^{2}$.}
\end{figure}

In analogy with the impact parameter dependent parton distribution of the
proton, we introduce the same for the photon. By 
taking a Fourier transform with respect to the transverse momentum
transfer $\Delta^\perp$ we get the GPDs in the transverse impact 
parameter space.
\be
q (x,b^\perp)&=&{1\over (2\pi)^2}\int d^2 \Delta^\perp 
e^{-i\Delta^\perp \cdot b^\perp} F^q \nonumber \\
&=&{1\over 2 \pi}\int \Delta d\Delta J_0(\Delta b) F^q ;
\ee
\be
\tilde q (x,b^\perp)&=&{1\over (2\pi)^2}\int d^2 \Delta^\perp
e^{-i\Delta^\perp \cdot b^\perp} \tilde F^q \nonumber \\
&=&{1\over 2 \pi}\int \Delta d\Delta J_0(\Delta b) \tilde F^q ;
\ee
where $J_0(z)$ is the Bessel function; $\Delta=|\Delta^\perp|$ and $b=|b^\perp|$.
In the numerical calculation, we have introduced a maximum limit 
$\Delta_{max}$ of the $\Delta$
integration which we restrict to satisfy the kinematics $-t<<Q^2$
\cite{hadron_optics,chiral,quark,model}. $q(x,b^\perp)$ gives the distribution
of partons in this case inside the photon  in the transverse plane. 
Like the proton, this interpretation holds in the
infinite momentum frame and there is no relativistic correction
to this identification because in light-front formalism, 
as well as in the infinite momentum
frame, the transverse boosts act like non-relativistic Galilean boosts.
$q(x, b^\perp)$ gives simultaneous information about the longitudinal
momentum fraction $x$ and the transverse distance $b$ of the parton from the
center of the photon and thus gives a new insight to the internal structure 
of the photon. The impact parameter distribution for a polarized photon is
given by $\tilde q(x, b^\perp)$. 

\begin{figure}[!htp]
%\centering
\begin{minipage}[c]{0.9\textwidth}

\tiny{(a)}\includegraphics[width=9.5cm,height=7cm,clip]{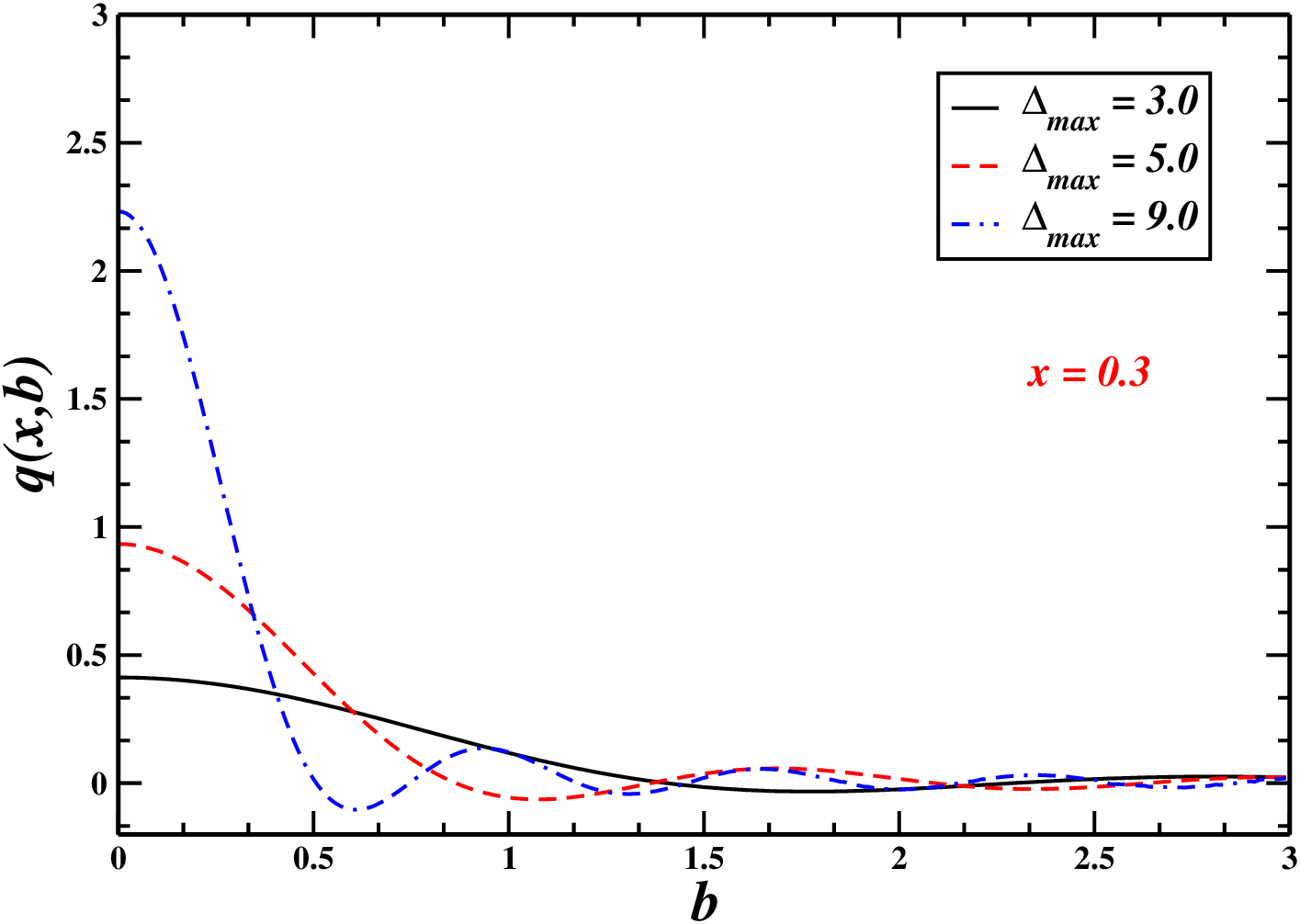}\\
\vspace{0.3in}

\tiny{(b)}\includegraphics[width=9.5cm,height=7cm,clip]{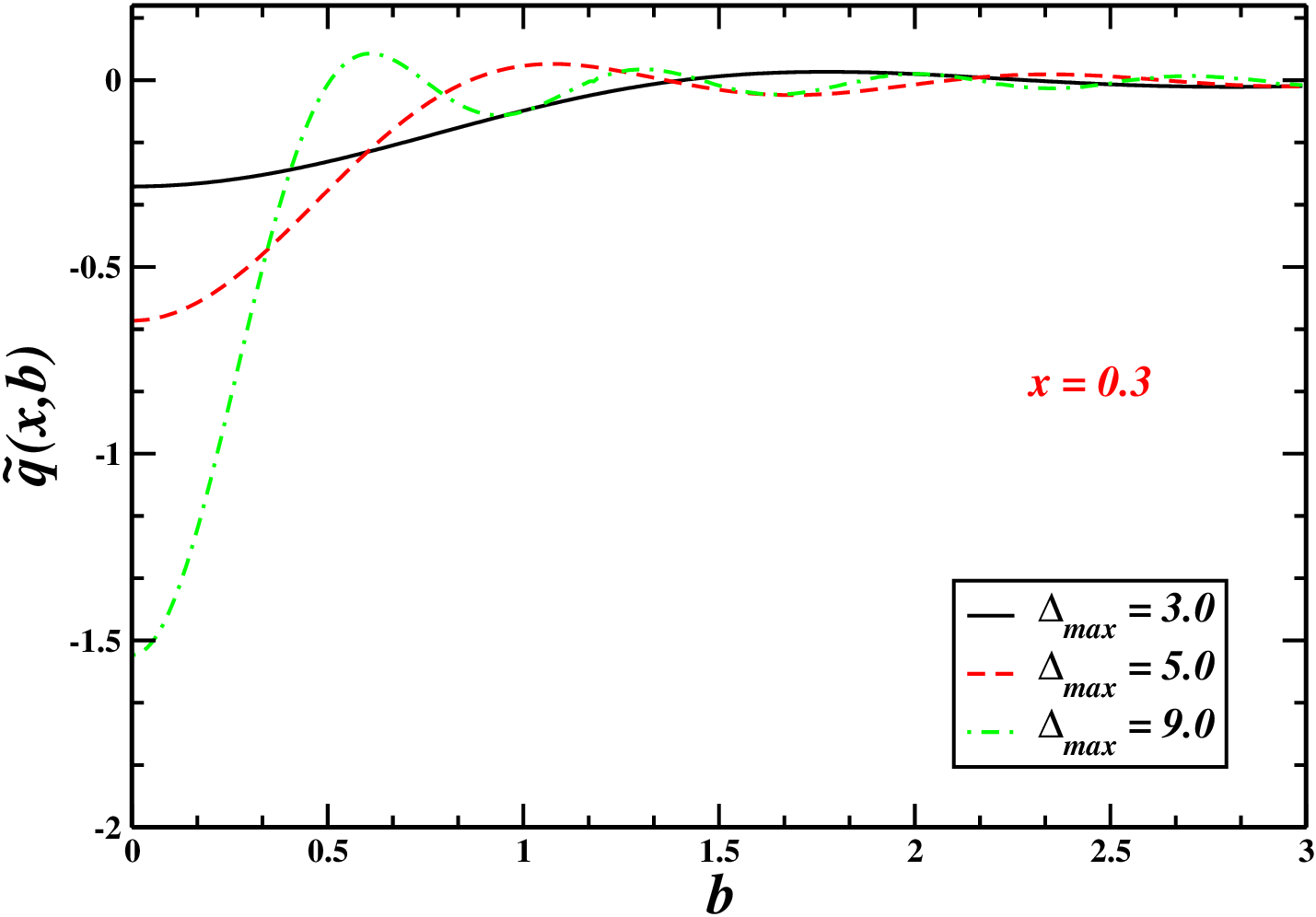}\\
\end{minipage}

\caption{\label{fig4}(Color online) Plots of impact parameter dependent
pdfs (a) ${q}(x,b)$  and (b) $\tilde{q}(x,b)$ vs $b$  for a fixed $x$;
where we have taken $\Lambda$ = 20 $\mathrm{GeV}$ for different values of 
$\Delta_{max}$ in GeV. $b$ is in ${\mathrm{GeV}}^{-1}$ and  $q(x,b)$ as well
as $\tilde{q}(x,b)$  are in ${\mathrm{GeV}}^{2}$.}
\end{figure}

\vspace{0.2in}

%%%%%%%%%%%%%%%%%%%%%%%%%%%%%%%%%%%%%%%%%%%%%%%%%%%%%%%%%%%%%%%%%%%%%%%%%%%%%%%%%%%%%%%%%%%%%
{\bf{Numerical results}}
\vspace{0.2in}
%%%%%%%%%%%%%%%%%%%%%%%%%%%%%%%%%%%%%%%%%%%%%%%%%%%%%%%%%%%%%%%%%%%%%%%%%%%%%%%%%%%%%%%%%%%%%

We have plotted the unpolarized GPD  $F^q$ and the polarized
GPD $\tilde F^q$ for the photon in Figs 1(a) and (b) respectively as
functions of $x$ and for different values of $t= -{(\Delta^\perp)}^2$. In
all plots we took the momentum transfer to be purely in the transverse
direction. We took $m=3.3$ MeV;  $\Lambda=Q = 20 \mathrm{GeV}$,
$\mu=0$ MeV  and $\Delta_{max}$= 3.0 GeV where $\Delta_{max}$ is the upper limit of
the $\Delta^\perp$  integration in the Fourier transform. It is to be noted
that the photon structure function $F_2^{\gamma}(x, Q^2)$ has been explored
over a wide kinematical range, namely $0.001 < x < 0.9$ and $1.9 < Q^2 <
780 \mathrm{GeV}^2$ \cite{buras}. However, here we restrict ourselves to
study the general features of the photon GPDs at a fixed scale rather than
the scale evolution. We have divided the GPDs by the normalization constant 
to compare with \cite{pire} in
the limit of zero $t$. Indeed they agree. As $-t$ increases, $F^q$
becomes more and more asymmetric with respect to $x=1/2$ : this asymmetry is 
prominent for lower values of $x$ which is expected as the $\Delta^\perp$ or 
$t$ dependence is associated with a $(1-x)^2$ factor (see the analytic 
expressions). The slope of the polarized GPD changes with increasing 
${(\Delta^\perp)}^2$.  Note that in \cite{pire} as well
as in the solid line in figs 1(a) and 1(b), the subleading mass terms are
not taken into account. It is to be
noted that in all plots we have  taken $0<x<1$ for which the contribution
comes from the active quark in the photon ($q {\bar q}$). 
As $x \rightarrow 1$, most of the momentum is carried by the
quark in the photon and the GPDs become independent of $t$.  
The Fourier transform (FT) of the unpolarized GPD $F^q$ is plotted in Fig. 2. 
Fig. 2 (a) shows the plot of the impact parameter dependent pdf of the photon as a
function of $x$ and for fixed impact parameter $b= \mid b^\perp \mid$. Fig.
2 (b) shows the same but as a function of $b$ and for fixed $x$.  The smearing in
$b^\perp$ space reveals the partonic substructure of the photon and its
'shape' in transverse space. In the ideal definition the Fourier transform
over $\Delta$ should be from $0$ to $\infty$. In this case the
$\Delta^\perp$ independent terms in $F^q$ and $\tilde{F}^q$ would give 
$\delta^2(b^\perp)$ in the  impact parameter space. This means in
the case of no transverse momentum transfer, the photon behaves like a point
particle in transverse position space. The distribution in transverse space 
is a unique feature accessible only when there is non-zero momentum transfer
in the transverse direction. From the plots it can be seen that for fixed 
$b$, $q(x,b)$ decreases slowly with $x$ till $x \approx
1/2$ then increases. The behavior in impact parameter space is
qualitatively different than a dressed quark target and also from 
phenomenological models of proton GPDs. For  a dressed quark target the
leading contribution to the GPD comes from the single particle sector of the
Fock space expansion, which in impact parameter space gives a delta
function peak. This contributes at $x=1$. For large values of $x$ the
peak in impact parameter space becomes sharper and narrower, that means
there is a higher probability of finding the active quark near the transverse center
of momentum \cite{quark}. In the phenomenological parametrization of proton GPDs where a
spectator model with Regge-type modification was used, the GPDs have a
different behavior in impact parameter space, the $u$ quark GPDs increase
with increasing $x$ for fixed $b$, reaches a maximum, then decrease. The peak
decreases with increasing $b$ \cite{model}. In the case of a photon there is 
no single particle contribution, and the distribution in $b$ space purely
reveals the internal $q {\bar q}$ structure  of the photon. Here near 
$ x \approx 1/2$ the peak in $b$ space is very broad which
means that the parton distribution is more dispersed when the $q$ and
${\bar q}$ share almost equal momenta. The parton distribution is sharper
both for smaller $x$ and larger $x$. In Figs 3 (a) and (b) we have plotted
the polarized distribution in impact parameter space. Fig. 3 (a)
shows it as a function of $x$ for fixed $b$ values and fig 3 (b) shows it as
a function of $b$ for fixed $x$ values. The slope decreases 
for higher $b$. The sign of the GPD changes at $x=1/2$, at which point the GPD and the pdf
in impact parameter space becomes zero. The distributions are approximately symmetric 
about $x=1/2$ in impact parameter space; as $x \rightarrow 1$ the
distribution increases sharply as the GPDs become independent of $t$; this
is similar to $q(x,b)$.  For fixed $x$, $\tilde q(x,b)$ as a function of $b$ 
becomes broader as $x$ increases until $x =1/2$. For larger values of $x$, 
it changes sign. In the plots we have 
taken the upper limit of the Fourier transform to be much smaller than $\Lambda$.
 The dependence of $q(x,b)$ and $\tilde{q}(x,b)$ on $\Delta_{max}$ is shown in
Figs. 4 (a) and (b) respectively. As $\Delta_{max}$ increases, the
distribution becomes sharper in impact parameter space. This shows that
larger momentum transfer probes the partons near the transverse center of the
photon. 

\vspace{0.2in}

%%%%%%%%%%%%%%%%%%%%%%%%%%%%%%%%%%%%%%%%%%%%%%%%%%%%%%%%%%%%%%%%%%%%%%%%%%%%%%%
{\bf{Conclusion}}
\vspace{0.2in}

%%%%%%%%%%%%%%%%%%%%%%%%%%%%%%%%%%%%%%%%%%%%%%%%%%%%%%%%%%%%%%%%%%%%%%%%%%%%%%
We presented a first calculation of the generalized parton distributions of
the photon, both polarized and unpolarized, when the momentum transfer in the
transverse direction is non-zero; at zeroth order in $\alpha_s$ and leading
order in $\alpha$; we calculated at leading logarithmic order and also kept
the mass terms at the vertex. We took the skewness to be zero. 
We express the GPDs in
terms of overlaps of the photon light-front wave functions. We considered the
matrix elements when the photon helicity is not flipped. When the momentum
transfer in the transverse direction is non-zero, one also has helcity flip
contributions, which will be treated in a later work. In our case only 
the diagonal parton number conserving overlaps contribute. We considered
both the quark and the antiquark contributions. The GPDs thus probe the two
particle $q {\bar q}$ structure of the photon. Taking a Fourier transform
(FT) with respect to $\Delta_\perp$ we obtain impact parameter dependent
parton distribution of the photon. We plot them for both polarized and
unpolarized photon. The parton distributions in impact parameter space show   
distinctive features compared to the proton and also compared to a 
dressed quark, which can be taken as an example of a spin $1/2$ composite
relativistic system consisting of a quark and a gluon. It is to be noted that
a complete understanding of the photon GPDs beyond leading logs would
require also the non-pointlike hadronic contributions which will be model
dependent \cite{pire}. However,  the GPDs of the
photon calculated here may act as interesting tools to understand 
the partonic substructure
of the photon. Accessing them in experiment is a challenge. On the
theoretical side, the next step would be to investigate the photon GPDs when
there is non-zero momentum transfer both in the transverse and in the
longitudinal direction as well as a perturbative study of the general properties of
GPDs like positivity and polynomiality conditions and sum rules.    

%%%%%%%%%%%%%%%%%%%%%%%%%%%%%%%%%%%%%%%%%%%%%%%%%%%
{\bf{Acknowledgments}}
\vspace{0.2in}

%%%%%%%%%%%%%%%%%%%%%%%%%%%%%%%%%%%%%%%%%%%%%%%%%%%%%% 
This work is supported by BRNS grant Sanction No. 2007/37/60/BRNS/2913 
dated 31.3.08, Govt. of India. We thank B. Pire for suggesting this topic
and B. Pire, S. Wallon and M. Diehl for helpful discussions. AM thanks Ecole
Polytechnique for support where part of this work was done.     

%%%%%%%%%%%%%%%%%%%%%%%%%%%%%%%%%%%%%%%%%%%%%%%%%%%%%%%%%%%%

\end{document}